\documentclass[11pt,twoside]{article}

\usepackage{asp2006}
\usepackage{epsf}
\usepackage{lscape}

\begin{document}
\title{Far-infrared all sky diffuse mapping with $AKARI$}   

\author{Y. Doi,$^1$ M. Etxaluze Azkonaga,$^{2,3}$ Glenn J. White,$^{2,3}$ E. Figueredo,$^4$ Y. Chinone,$^5$ M. Hattori,$^5$ T. Nakagawa,$^6$ C. Yamauchi,$^6$ H. Shibai,$^7$ and the $AKARI$ Diffuse Map team}   
\affil{
$^1$ Department of Earth Science and Astronomy, 
The University of Tokyo, Japan\\
$^2$ Department of Physics \& Astronomy, The Open University, UK \\
$^3$ Space Science and Technology Dept., 
 Rutherford Appleton Laboratory, UK\\
$^4$ Instituto de Astronomia, 
Universidade de S\~ao Paulo, Brazil\\
$^5$ Astronomical Institute, Tohoku University, Japan\\
$^6$ Institute of Space and Astronautical Science, JAXA, Japan \\
$^7$ Department of Earth and Space Science, 
Osaka University, Japan
}    

\begin{abstract} 
We discuss the capability of $AKARI$ in recovering diffuse far-infrared emission, and examine
 the achieved reliability.
Critical issues in making images of diffuse emission are the transient response and long-term stability of the far-infrared detectors. Quantitative evaluation of these characteristics are the key to achieving sensitivity comparable to or better than that for point sources ($< 20$ -- $95$ MJy sr$^{-1}$). We describe current activity and progress toward the production of high quality images of the diffuse far-infrared emission using the $AKARI$ all-sky survey data.
\end{abstract}
\vspace*{-5mm}



\vspace*{-3mm}
\section{$AKARI$ far-infrared All-Sky Survey}\label{sec:allsky}

\begin{figure}[t]
\begin{minipage}[b]{0.42\linewidth}
   \centering
   \resizebox{0.95\hsize}{!}{
      \includegraphics*{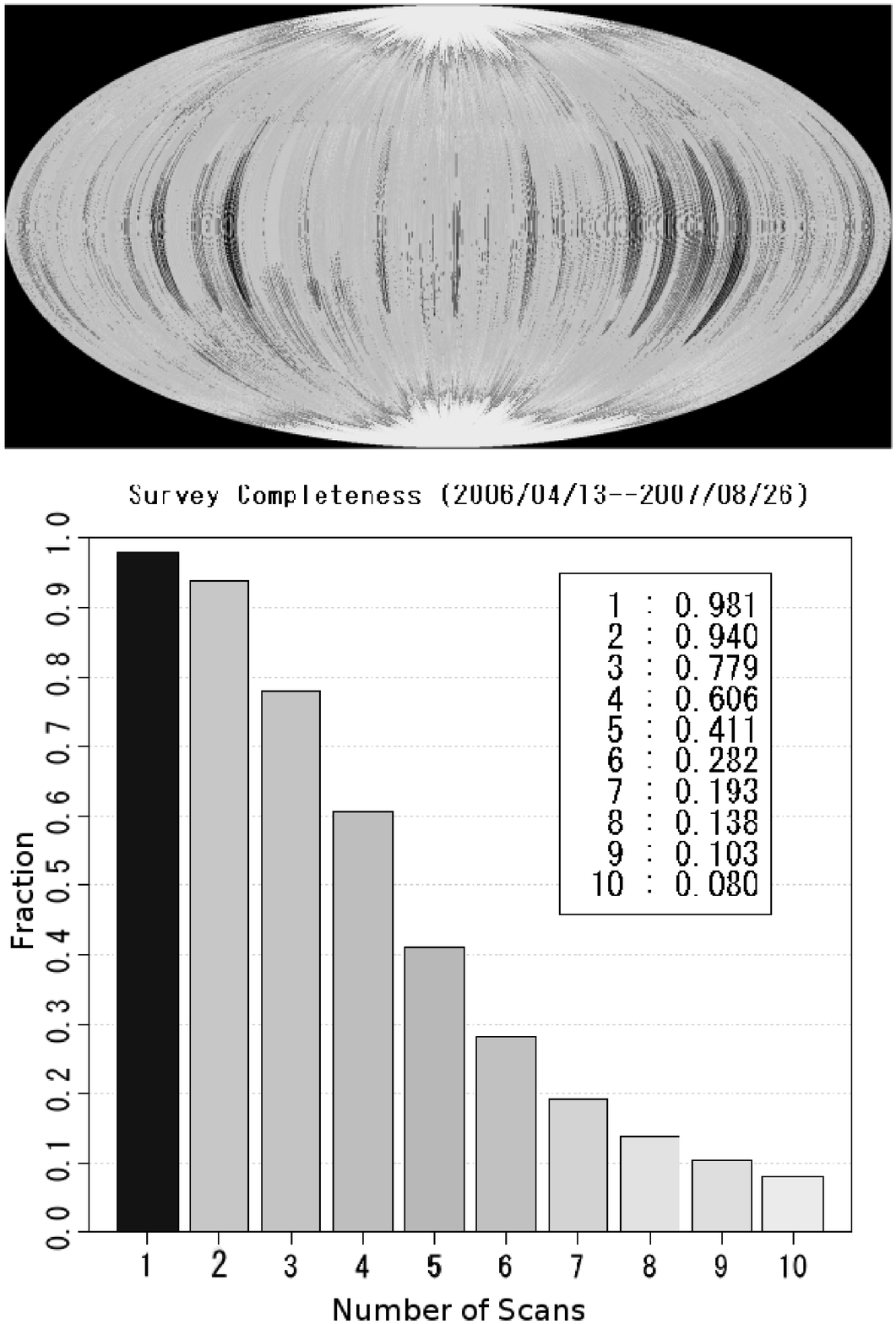}
   }
   \caption{
     Sky coverage of the $AKARI$ far-infrared all-sky survey displayed in ecliptic coordinates (upper panel).
     Lower panel shows histogram of number of observational scans.
   }\label{fig:skycov}
\vspace*{-3mm}
\end{minipage}
\begin{minipage}[b]{0.58\linewidth}
   \centering
   \resizebox{0.95\hsize}{!}{
      \includegraphics*{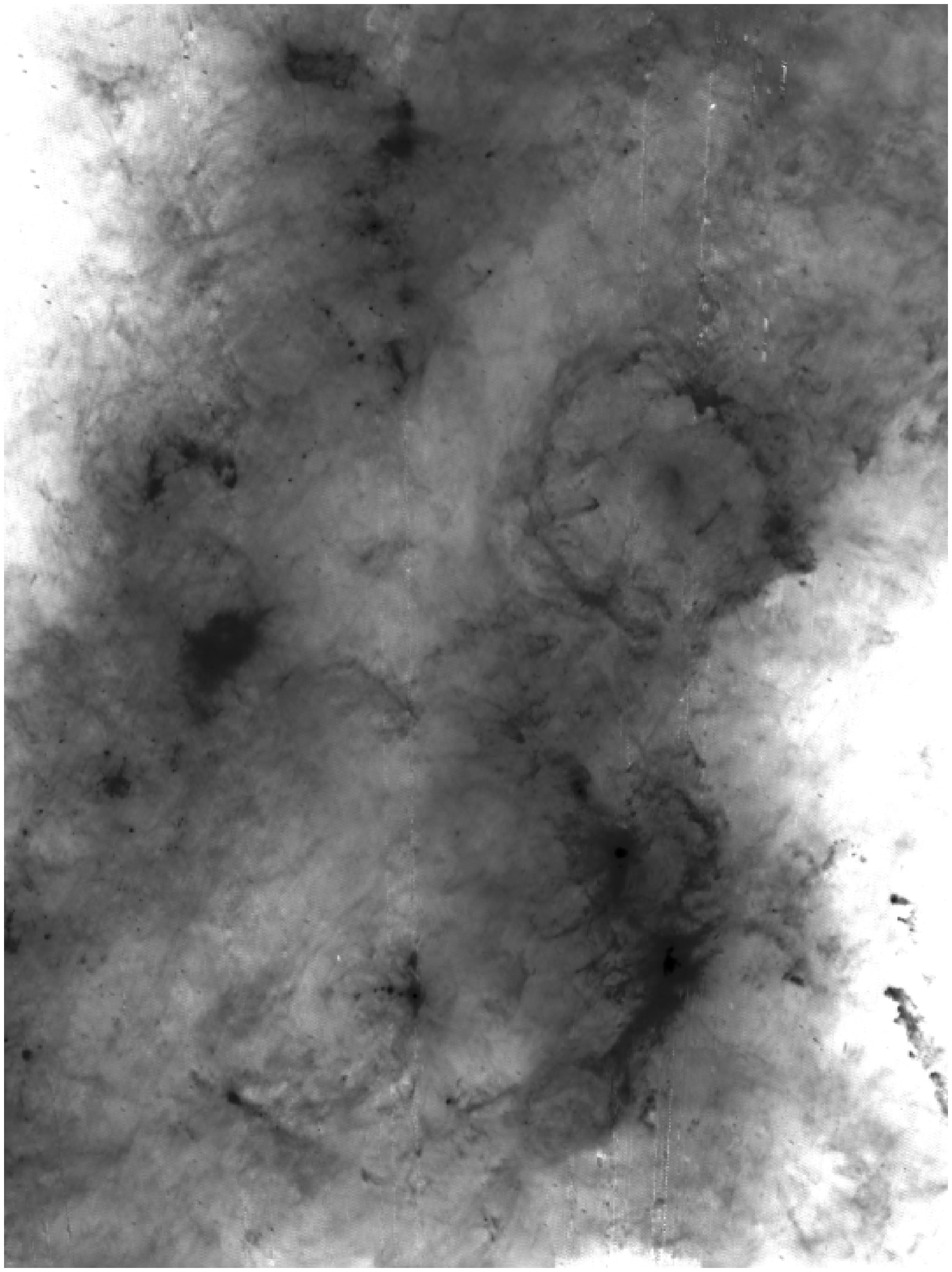}
   }
   \caption{
     $AKARI$ 140 $\mu$m continuum image of the Orion region  $(20^{\circ}\times40^{\circ})$.
   }\label{fig:orion}
\vspace*{-3mm}
\end{minipage}
\end{figure}

The $AKARI$ infrared all-sky survey is the direct successor to the IRAS survey that was launched more than twenty five years ago.
$AKARI$'s sun-synchronous orbit permits the whole sky to be repeatedly surveyed every six months, and during the cold phase of the mission from 2006 April -- 2007 August, more than 94\% of the whole sky was observed (Figure~\ref{fig:skycov}). This survey covered the wavelength region $50\ \mu$m -- $180\ \mu$m with four spectral bands centered at $65~\mu$m, $90~\mu$m, $140~\mu$m, and $160~\mu$m, and achieved a spatial resolution of $43''$ -- $72''$, and detection limit from 0.6 -- 6 Jy (single scan, $5\sigma$).

\begin{figure}[bt]
\begin{minipage}[b]{0.50\linewidth}
   \centering
   \resizebox{1.0\hsize}{!}{
      \includegraphics*{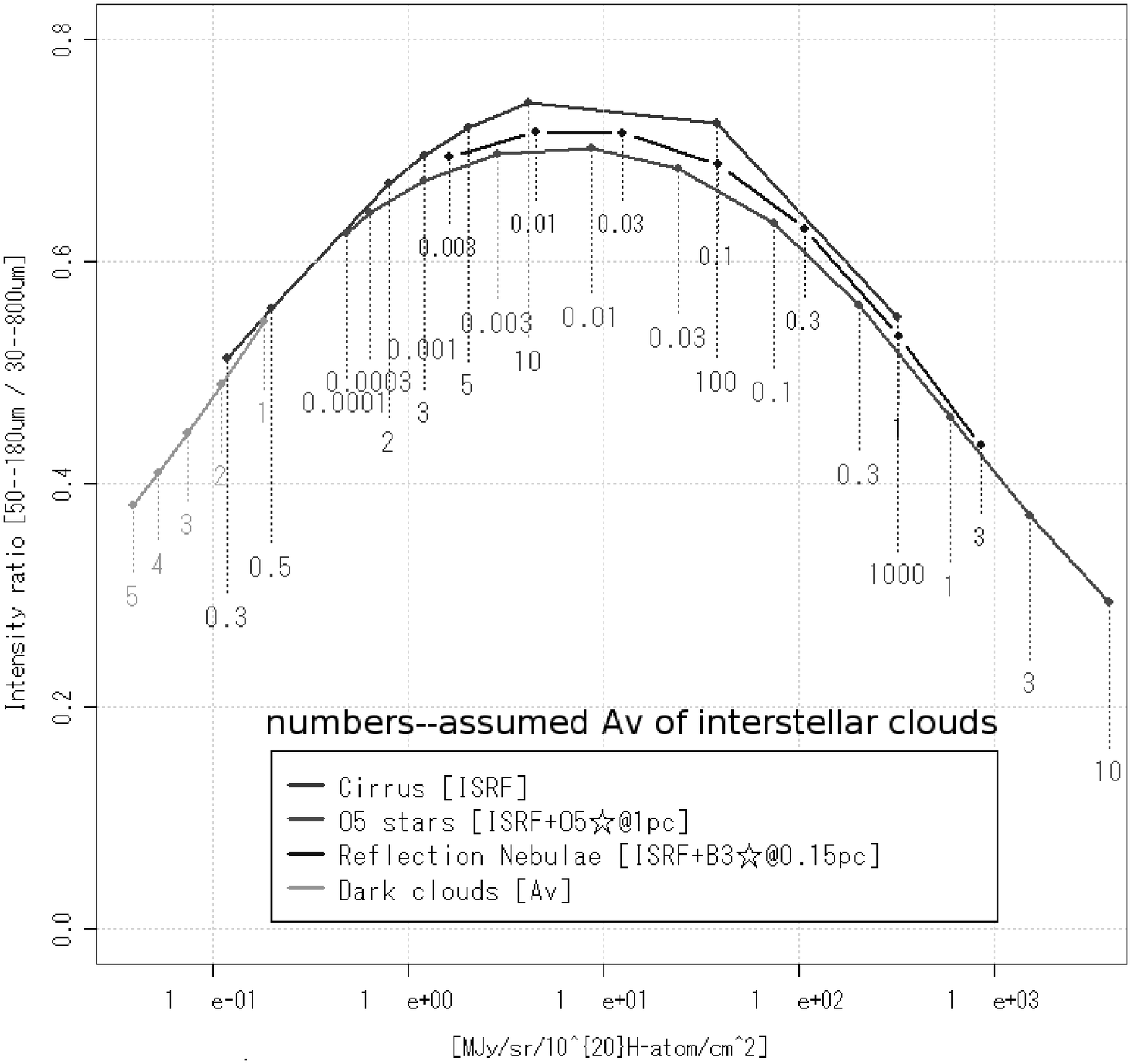}
   }
\end{minipage}
\begin{minipage}[b]{0.5\linewidth}
   \centering
   \resizebox{1.0\hsize}{!}{
     \includegraphics*{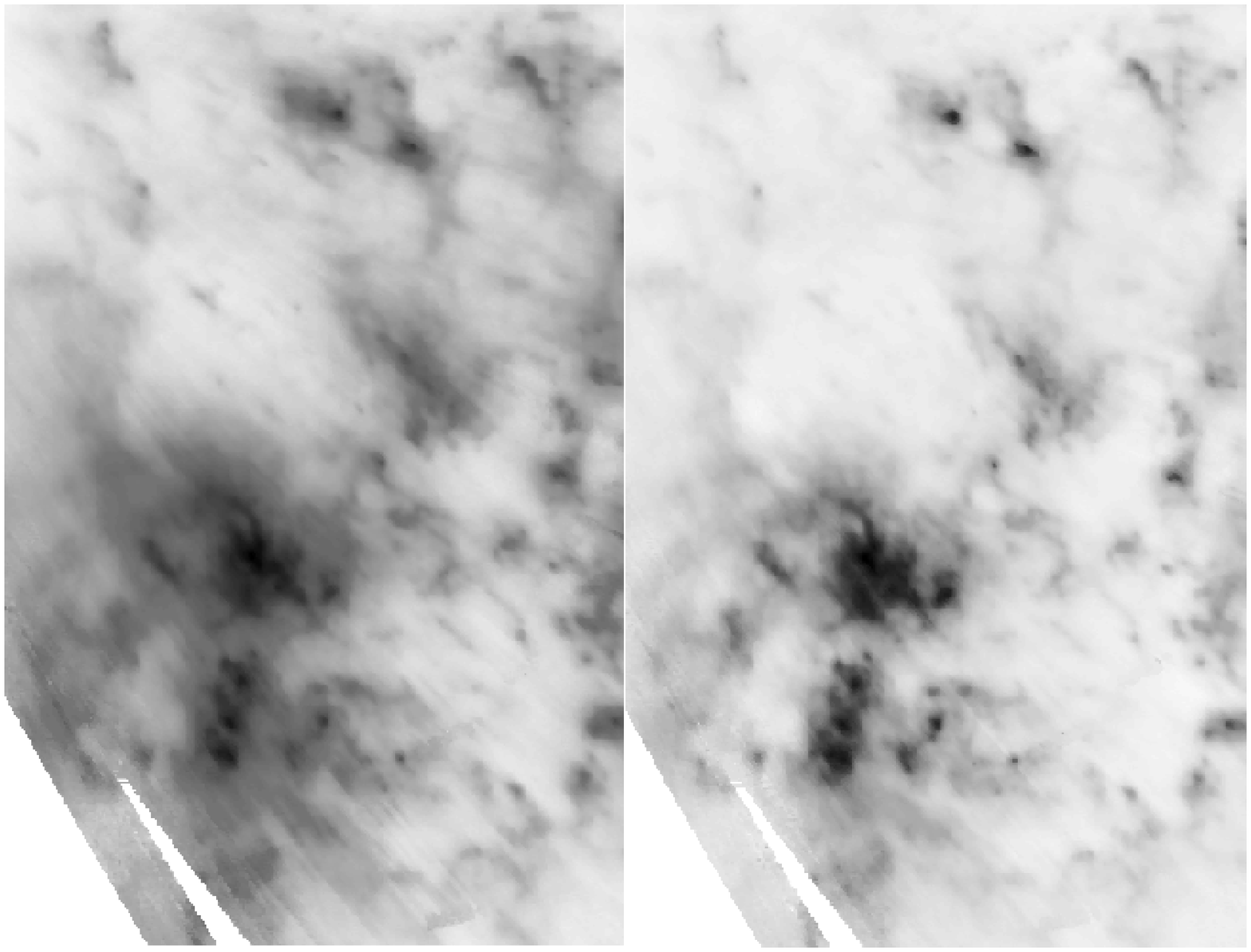}
   }
\end{minipage}
   \caption{
     Left panel -- 
estimated fraction of far-infrared continuum intensity observed in the $AKARI$ waveband (50--180 $\mu$m) over the total far-infrared intensity ($I_{\rm FIR}$) based on a model calculation by \citet{desert90} for interstellar matter under various conditions.
}\label{fig:ifir}
\vspace*{-.8mm}
\caption{
  Right panels -- effect of the slow-response correction. Image without correction (middle panel) and that with correction (right panel) are shown.
}\label{fig:srcorrection}
\vspace*{-3mm}
\end{figure}

The far-infrared emission from dust grains is a diagnostic of the total energy output of newly formed stars, providing a good indicator to the star-forming activity  \citep{kennicutt98}. The spectral coverage of the $AKARI$ detectors
 continuously
 covers the main portion of far-infrared continuum emission by interstellar dust particles, enabling us to make a precise estimation of the total far-infrared intensity and the luminosity ($I_{\rm FIR}$; see Figure~\ref{fig:ifir}).

\vspace*{-3mm}
\section{Data Analysis for Image Mapping}

The far-IR data were pre-processed using the $AKARI$ pipeline tool originally optimised for point source extraction \citep{yamamura09}. This included corrections for linearity and sensitivity drifts of the detectors, rejection of bad-data due to cosmic ray hits, and other instrumental effects. Following this initial pre-processing, a separate pipeline is then used to recover the large-scale spatial structures accurately. The Diffuse mapping pipeline additionally corrects for the transient response of the far-infrared photoconductor detectors, which is found to be particularly important for the recovery of high quality diffuse structure \citep[][Figure~\ref{fig:srcorrection}]{shirahata09}. \citet{kaneda09} have previously discussed the characterisation of detector slow-response effects and their mitigation, and we take a practical approach here to correct the all-sky survey data so that high quality diffuse maps can be recovered with realistic computational overheads.

\begin{figure}[t]
\begin{minipage}[b]{0.4\linewidth}
   \centering
   \resizebox{0.95\hsize}{!}{
     \includegraphics*{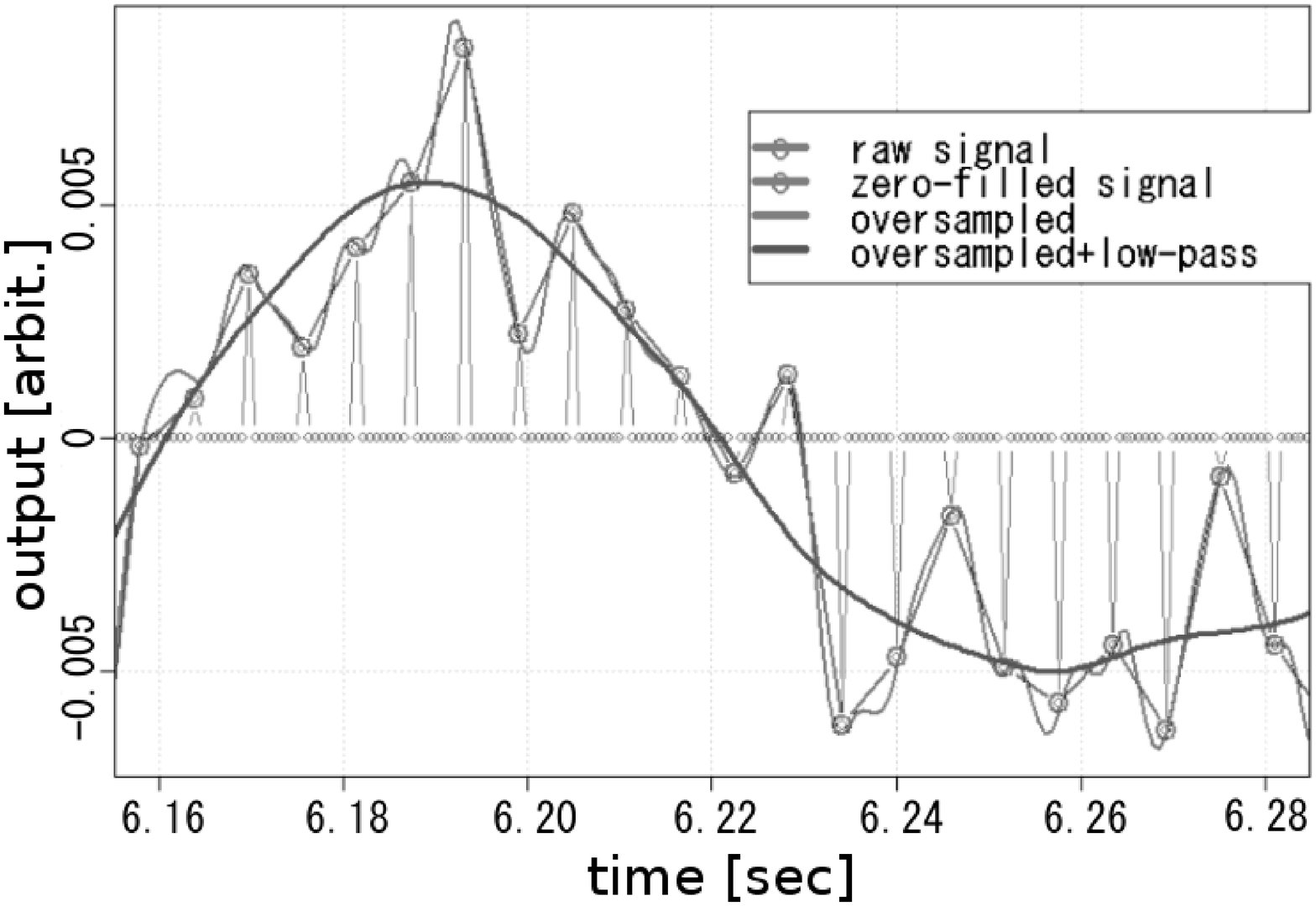}
   }
\end{minipage}
\begin{minipage}[b]{0.6\linewidth}
   \centering
   \resizebox{0.95\hsize}{!}{
     \includegraphics*{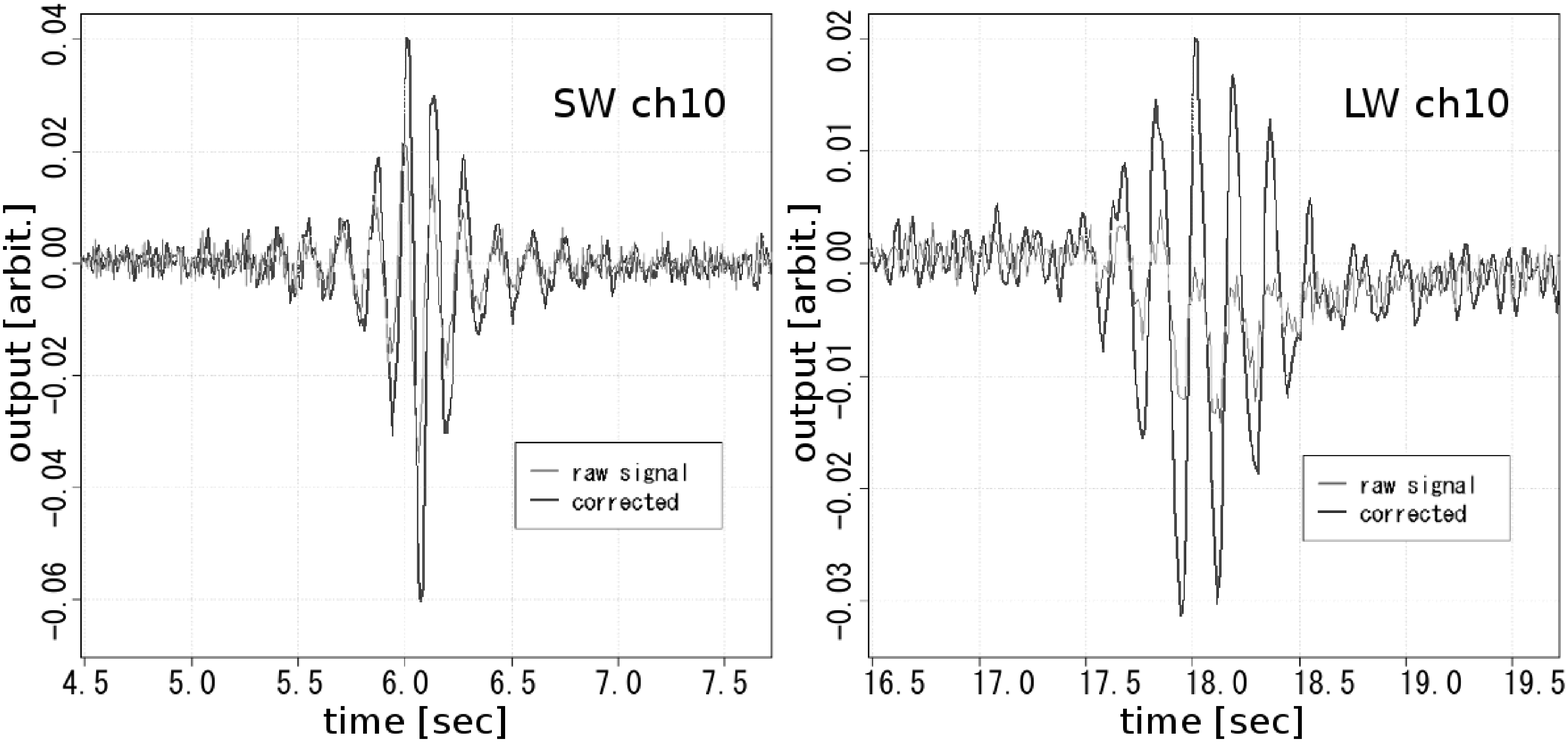}
   }
\end{minipage}
\caption{
  Left panel -- oversampling of the input signal by inserting equally spaced null data between the original data.
Low-pass filtering the data results smoothly interpolated data conserving its original frequency information and have higher sampling rate.
Here we apply a low-pass filter with a cut-off frequency that is lower than original sampling rate, so that we can eliminate high-frequency noise contaminating the original data simultaneously.
}\label{fig:oversample}
\caption{
  Right panels -- examples of the slow-response correction applied to FTS zero-pass signal (see text).
  The left panel shows data from the SW channel and the right panel shows that of the LW channel.
  Raw signals that are heavily distorted by
 the slow-response of the detector are restored to plausible symmetrical signals by the correction.
  Note that high-frequency noise seen in the raw signals is smoothed out in the processed signals by low-pass filtering applied concurrently in the correction process (see Figure~\ref{fig:oversample}).
}\label{fig:FTScorrection}
\vspace*{-3mm}
\end{figure}

We evaluated the slow step response of the detector signal using an internal calibration light. The detector frequency response can then be evaluated from the step response to the lamp by F\'ourier transformation. The slow-response correction can be made by filtering the raw signal using a numerical filter whose frequency response is the inverse of that of the detector. We apply an infinite impulse response filter (IIR filter) to make this correction with high precision, which is computationally much more efficient than, e.g., using a finite impulse response filter (FIR filter). Oversampling the detector signal by a factor of eight was also adopted (Figure~\ref{fig:oversample}) to better approximate the filter's frequency response.

Using this procedure, we were able to process a one-hour section of all-sky survey data in each detector channel (43.3 min for SW data and 64.8 min for LW data) in 0.78 sec, which is an acceptable computing overhead that will allow processing the complete 1.5 years of survey data.

The reliability of the slow-response correction can be demonstrated by taking the FIS spectroscopic signal as a trial signal, since it should be symmetric around the zero-pass length of F\'ourier spectrometer optics \citep{kawada07}, and appears to yield plausible corrected signals (Figure~\ref{fig:FTScorrection}).

We further tested the method by comparing two data sets observing the same position in the sky but in the opposite spatial scan direction, taken half-a-year apart using the opportunities provided by the $AKARI$ satellite's sun-synchronous orbit ($\S$\ref{sec:allsky}).

\begin{figure}[t]
\begin{minipage}[b]{0.61\linewidth}
   \centering
   \resizebox{0.95\hsize}{!}{
      \includegraphics*{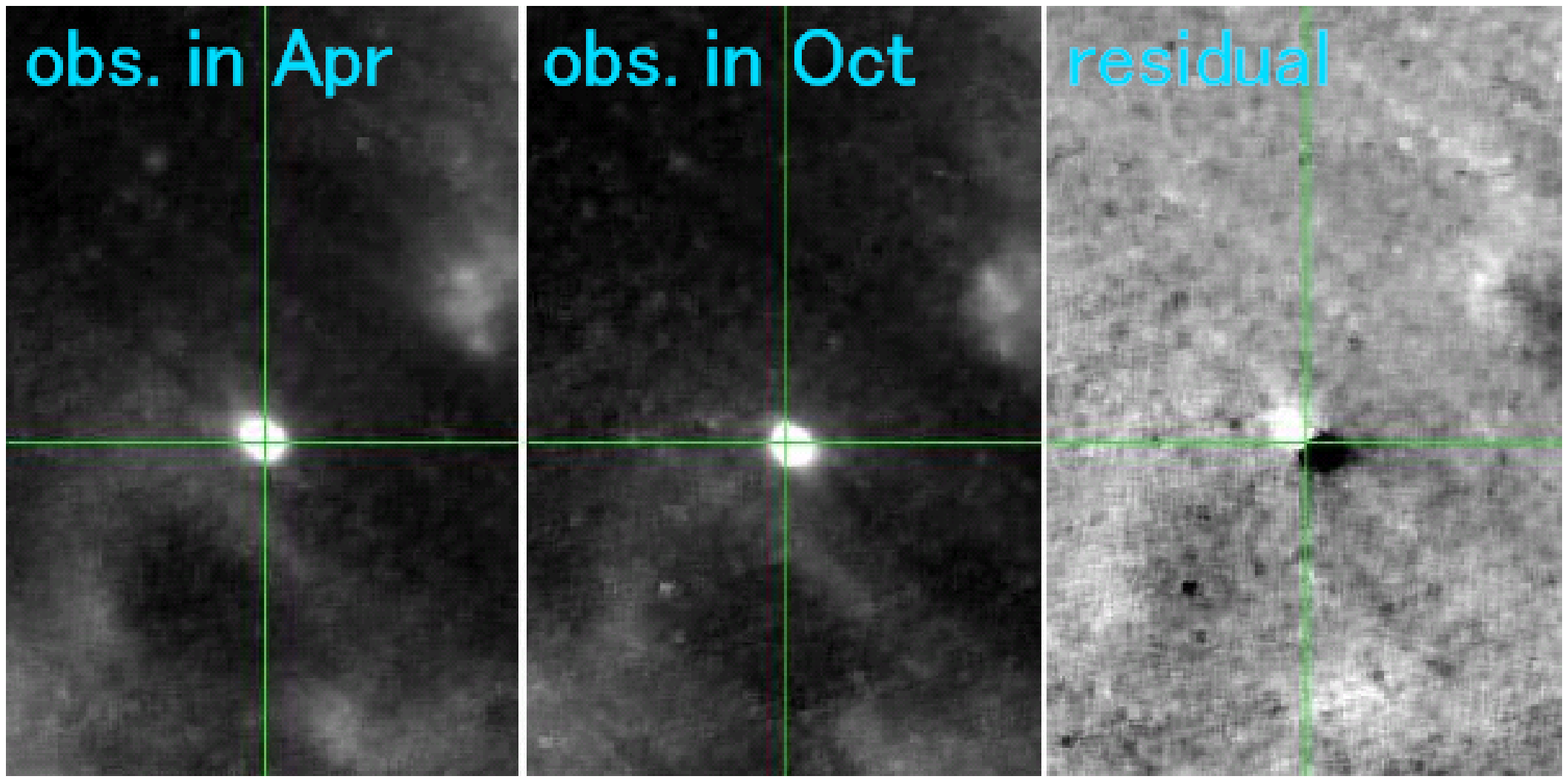}
   }
\end{minipage}
\begin{minipage}[b]{0.38\linewidth}
   \centering
   \resizebox{0.95\hsize}{!}{
      \includegraphics*{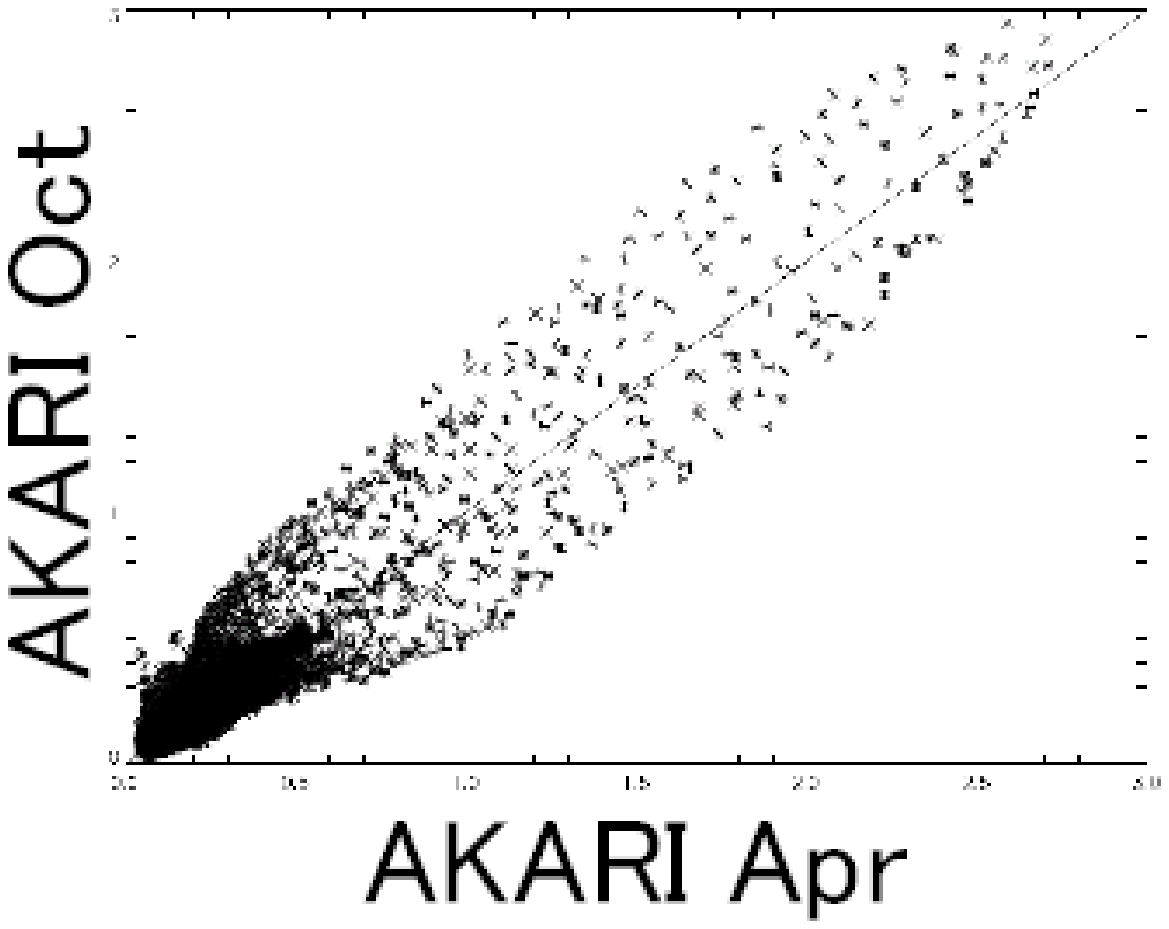}
   }
\end{minipage}
   \caption{
     Left panels -- comparison between two images observing same region in the sky but in the opposite spatial scan direction.
     The left panel shows an image taken in Apr. 2006 scanned from lower right to upper left, while the middle panel shows an image taken in Oct. 2006 scanned from upper left to lower right.
     The right panel shows difference between two images so that we could check the accordance of the two images.
   }\label{fig:difimage}
   \caption{
     Right panel -- a scatter plot showing the correlation between the two data sets shown in Figure~\ref{fig:difimage}.
     Larger scatter is found for data taken in the vicinity of a point source. See text for the detailed explanation.
   }\label{fig:correlation}
\vspace*{-3mm}
\end{figure}

An example of the comparison is shown in Figure~\ref{fig:difimage}. The data taken in April 2006 and in October 2006 show good correlation, although there are significant differences seen in the differentiated image due to the after effects of point source crossing. This deviation from the correlation can also be seen in a scatter plot of the same data (Figure~\ref{fig:correlation}). Sky locations having small far-IR intensities show good correlation while those having larger intensities, or those in the vicinity of point sources, show a larger scatter. We are currently investigating the precise correction of the detector slow-response for signals with various sky positions.

\vspace*{-3mm}
\begin{thebibliography}{}
\bibitem[D\'esert et al.(1990)]{desert90}
   D\'esert, F.~-X., Boulanger, F., \& Puget, J.~L., et al. 1990, \aap, 237, 215
\bibitem[Kaneda et al.(2009)]{kaneda09}
   Kaneda, H., Fouks, B., Yasuda, A., et al. 2009, \pasp, in press
\bibitem[Kawada et al.(2007)]{kawada07}
   Kawada, M., Baba, H., Barthel, P.~D., et al. 2007, \pasj, 59, S389
\bibitem[Kennicutt (1998)]{kennicutt98}
   Kennicutt, Jr., R.~C. 1998, \araa, 36, 189
\bibitem[Shirahata et al.(2009)]{shirahata09}
   Shirahata, M., Matsuura, S., Hasegawa, S., et al.  2009, \pasj, in press
\bibitem[Yamamura et al.(2009)]{yamamura09}
   Yamamura, I., Makiuti, S., Ikeda, N., et al. 2009, this volume

\end{thebibliography}
\end{document}